\documentclass[runningheads]{llncs}

\usepackage{hyperref}

\begin{document}
%\titlerunning{Improved vectorization of OpenCV algorithms for RISC-V CPUs}

\title{Improved vectorization of OpenCV algorithms for RISC-V CPUs\thanks{The project is supported by the Lobachevsky University Academic Leadership Program ``Priority-2030''}}

\author{V.~D.~Volokitin\inst{1} \and
	E.~P.~Vasiliev\inst{1} \and
	E.~A.~Kozinov\inst{1} \and
	V.~D.~Kustikova\inst{1} \and
	A.~V.~Liniov\inst{1} \and
	Y.~A.~Rodimkov\inst{1} \and
	A.~V.~Sysoyev\inst{1} \and
	I.~B.~Meyerov\inst{1}}

\authorrunning{V.~D.~Volokitin et al.}
\institute{Lobachevsky State University of Nizhni Novgorod, Gagarin Avenue 23, Nizhni Novgorod, 603022, Russia
	\email{meerov@vmk.unn.ru}}
\maketitle              % typeset the header of the contribution
\begin{abstract}
The development of an open and free RISC-V architecture is of great interest for a wide range of areas, including high-performance computing and numerical simulation in mathematics, physics, chemistry and other problem domains. In this paper, we discuss the possibilities of accelerating computations on available RISC-V processors by improving the vectorization of several computer vision and machine learning algorithms in the widely used OpenCV library. It is shown that improved vectorization speeds up computations on existing prototypes of RISC-V devices by tens of percent.
	
\keywords{RISC-V \and OpenCV \and vectorization \and performance optimization \and high-performance computing.}
\end{abstract}

\section{Introduction}
OpenCV \cite{c1-1,c1-2} is a widely used open-source library of computer vision and machine learning algorithms. The library contains high-performance implementations of over 2,500 algorithms and is the de facto standard in video and image analysis. OpenCV is used in many application areas, including computational mathematics, physics, chemistry, biology, and other problem domains. For example, the methods of computer vision, machine learning, and image analysis developed in the library can be used to visualize the results of numerical simulations \cite{c2}, plan and analyze physical experiments \cite{c3}, and compare the results of numerical simulations with real experiments \cite{c4}. Many computer vision and machine learning algorithms require large computational resources, and their application in numerical simulation requires high-performance computers or even supercomputers. However, the currently used x86 and ARM architectures, with all their advantages, are not free from drawbacks. For example, the performance of the most powerful supercomputers in the HPCG test, which is based on memory-bound algorithms, is almost two orders of magnitude inferior to the performance in the LINPACK test, which is based on compute-bound algorithms. Moreover, processor architectures that are widely used today are proprietary and closed. Therefore, third-party developers cannot make changes to them and implement co-design ideas. Due to these factors, the development of new architectures is still of considerable interest.

Proposed about 12 years ago at the University of California at Berkeley, the free and open RISC-V architecture \cite{c5} is developing very rapidly. Over the years, RISC-V developers have gone from the proof-of-concept stage to high-performance RISC-V devices that are already available to purchase and test \cite{c6,c9}. It is noteworthy that the development of technologies and related software in a certain sense goes in a spiral. When several authors of this paper took part in the installation and testing of the first supercomputer at Lobachevsky University, state-of-the-art (at the beginning of the 2000s) Intel server CPUs already had powerful vector instruction sets, but the capabilities of compilers were extremely limited compared to the current state of development. For example, when utilizing compiler autovectorization for a simple loop that calculates the sum of an array, it was necessary to manually unroll the loop to prevent a data dependency that could be easily resolved automatically by the compiler. However, with each subsequent version of the Intel Compiler and GCC, significant progress was observed, and many optimization techniques became unnecessary and only led to code clutter. State-of-the-art versions of compilers for x86 CPUs have advanced vectorization capabilities and in many cases do an excellent job of translating the code into SIMD instructions. Performance models implemented in the compilers make it possible not only to vectorize the code, but often to choose the best available option for the CPU architecture, taking into account its unique features. In case of vectorization inefficiency, an optimizing compiler often warns the developer in the optimization report and helps to avoid performance degradation. The compilers for x86 architectures have not only advanced autovectorization capabilities, but also powerful code fine-tuning tools for performance engineers. Compilers are complemented by performance analysis and optimization tools such as, for example, Intel VTune \cite{c11,c12}, Intel Advisor \cite{c13,c14,c15}, and likwid \cite{c10}. 

The available compilers for RISC-V CPUs are at the beginning of this path, so achieving acceptable performance on RISC-V devices requires some effort from the programmer. In this paper, we continue the project started in \cite{c6} and present our developments in the field of performance analysis and optimization of several OpenCV algorithms for multicore RISC-V processors with SIMD extensions. We compare and analyze the performance of available RISC-V devices using image filtering, morphology transformations (erosion), and bag-of-words algorithms as a testbed, and demonstrate how to improve vectorization taking into account the properties of existing RISC-V CPUs. The results are publicly available \cite{c16} and can be used in research and development.

\section{OpenCV Overview}

\subsection{What is OpenCV?}

OpenCV is an open-source library of computer vision algorithms distributed under the Apache 2.0 license. The library can be used for both research and commercial purposes. OpenCV has been in development for over 20 years and has proven itself in the user community. The library is actively used for rapid prototyping by developers of video surveillance and image analysis software since it contains the implementation of a large number of state-of-the-art computer vision algorithms and provides opportunities for creating a complete application pipeline from data preprocessing to displaying the results. OpenCV is used to solve problems of image understanding (image classification, object detection and recognition, image segmentation, instance segmentation) and video analysis (tracking objects, motion estimation), augmented reality problems, mobile robotics, and many others. The library also contains implementations of a large number of basic machine learning algorithms. 

The library is developed in the C++ programming language and has wrappers for Python, Java, and MATLAB/Octave. In later versions, it contains JavaScript bindings for a certain set of functions. OpenCV is a cross-platform library that works under Linux, Windows, macOS, Android, iOS, and some other operating systems. OpenCV has a modular structure. When building a library, the package includes shared or static libraries that can be used in application development. The main modules are as follows \cite{c1-1,c7}:

\begin{enumerate}
  \item \textit{Base module (core).} Includes the declaration of basic data types and functions for data processing. 
  \item \textit{Image processing module (imgproc).} Contains, in particular, the implementation of various filtering operations, geometric transformations (scaling, affine and perspective deformations, and others), and color space transformations. 
  \item \textit{Video analysis module (video).} Includes the implementation of motion estimation and background subtraction algorithms, as well as various methods for tracking objects in video frames. 
  \item \textit{Module of algorithms for camera calibration and 3D reconstruction (calib3d).} 
  \item \textit{Module for searching for two-dimensional features (features2d).} Contains the implementation of detectors and descriptors of keypoints on images, as well as algorithms for matching keypoints on a pair of images. 
  \item \textit{A module that provides a high-level interface for building a simple graphical interface (highgui).} 
  \item \textit{Video input-output module (videoio).} 
  \item \textit{Machine learning module (ml).} Includes the implementation of many machine learning algorithms, namely boosting, decision trees, gradient boosting trees, random forest, support vector machine, expectation-maximization algorithm, and several others. 
  \item \textit{Module for working with deep neural network models (dnn).} Provides inference of deep neural networks trained using third-party frameworks (Caffe, PyTorch, TensorFlow). 
  \item Other \textit{auxiliary modules} containing implementations of commonly used problem-specific algorithms, as well as modules that ensure efficient execution on specific computing devices.
\end{enumerate}

\subsection{Vectorization and Parallelization in OpenCV}

Successful vectorization of algorithms is the key to achieving competitive performance and efficient use of the computing resources of modern CPUs. In this regard, the OpenCV developers implemented so-called ``universal intrinsics'' to make the code portable and to employ vector processing units efficiently. Universal intrinsics are a high-level abstraction for vectorization methods. They allow the use of built-in functions without developing and maintaining platform-specific codes. OpenCV supports various types of 128-bit registers for a wide range of architectures, including x86 (SSE/SSE2/SSE4.2), ARM (NEON), PowerPC (VSX), MIPS (MCA), RISC-V (RVV~0.7.1). 256-bit and 512-bit registers are supported on x86 processors (AVX2 and AVX-512 instruction set extensions, respectively) \cite{c1-1,c8}.

Parallelization of algorithms in OpenCV is provided by introducing the \texttt{parallel\_for\_} abstraction, which is implemented using OpenMP, TBB, or pthreads, depending on which framework is explicitly specified or detected during the OpenCV build. It's worth noting that a sequential version of the library can also be built. The CMake utility is used to simplify the building.

\section{Improving Vectorization of OpenCV Algorithms for RISC-V CPUs}

As mentioned above, the OpenCV library uses the implementation of vector algorithms through universal intrinsics \cite{c8}. This method allows abstracting from a specific architecture and implements a portable vector code using specific functions and data types focused on working with vectors. This approach has obvious merits. Thus, after writing the vector implementation of the algorithm, there is no need to rewrite the code when a new set of vector instructions appears in the hardware. Of course, this approach only works if the library has an implementation of universal intrinsics for the target architecture. Note that in this case, the vectorized code works \textit{without any overhead}, being translated into native vector instructions specific to a particular architecture. The disadvantage of this approach should also be noted. Sometimes, such a ``universal'' code will not be translated into the most optimal sequence of instructions, since it is advisable to develop code taking into account certain features of a target architecture. Nevertheless, this consideration does not change the general assessment of the approach that provides portability and independence from the architecture, which is the main and obvious achievement. In this paper, we demonstrate that using this approach it is possible to significantly improve the performance of several OpenCV algorithms with relatively small code modifications while maintaining its portability.

Given our interest in the vector implementation of algorithms for the RISC-V architecture, we will focus on the implementation of the universal OpenCV intrinsics for RISC-V. OpenCV currently supports RVV~0.7.1 which is a 128-bit vector instruction set for RISC-V CPUs. The RVV vector instruction sets are supported in some currently available RISC-V processors. The computational cores developed by T-Head implement the RVV~0.7.1 vector extension. Note that the official version of the RVV~1.0 extension exists, but at present it is implemented in single copies only. The RVV~0.7.1 extension is used in a large number of devices, which allows us to try it on available hardware right now. The two versions of these vector standards do not provide for backward compatibility, but there is specialized software \cite{c9} that allows translating RVV~1.0 commands into RVV~0.7.1 commands at the assembler level. The RVV~0.7.1 extension provides for working not only with one vector register but also with a consecutive block of such registers (1, 2, 4, or 8 registers). This is very similar to the presence of ``extended'' registers, which allow CPU to process a larger amount of data and further speed up calculations. Let us consider in more detail the reasons for the possible acceleration as a result of the transition to a wider vector length. First, we note that an ``extended'' (for example, 512-bit) instruction at the architecture (assembler code) level can be decomposed into a sequence of 128-bit micro-operations in the microarchitecture. For example, using 512-bit instructions in a loop instead of 128-bit ones is similar to unrolling the loop for 4 iterations. Compared to the original code without loop unrolling, this can result in a speedup by reducing the number of control operations and the need to predict and execute conditional jumps. Secondly, when using this approach, the work on the decoding stage is reduced. Thirdly, in some cases, when using extended vector instructions, we save the resources of the memory subsystem. In general, this approach can lead to a significant gain in code performance, which will be shown below.

In this project, we have made changes to the implementation of the universal intrinsics of OpenCV so that they are compiled into extended 512-bit RISC-V vector instructions where possible. It usually makes sense to increase the size of vector registers (to be more precise, the size of the block of registers), if the algorithm allows it, and the memory subsystem has time to load data. However, in this case, we chose a block size of 512 bits (4 registers) because some OpenCV functions return results in extended precision. Then, working with a block of 4 registers, we will end up with a block of 8 registers, which is the maximum size of a block of registers in the available RISC-V CPUs. Below is a brief description of the changes made to implement the universal OpenCV intrinsics for the RISC-V architecture. 

The structure of the intrinsics provided by the T-Head team \cite{c20} in the GCC~8.4 compiler is as follows: \texttt{<command name>\_<data set>m<number of registers>}.

For example, the FMA instruction between vectors for the float data type is as follows: \texttt{vfmadd\_vv\_f32m1} for a 1-register vector and \texttt{vfmadd\_vv\_f32m4} for a 4-register vector. The data types employed by these commands differ similarly (in the number of registers used). Based on this, the main changes in the OpenCV code were the implementation of the OpenCV intrinsics through commands and data types that use 4 registers, and not 1, as is now implemented in the main branch of OpenCV \cite{c17-2}. Also, for correct operation, it was necessary to add several functions for transforming wide vectors, since not all of them are present and implemented in the current compiler (for example, we need transformations from m8 to m4 data type, which must be implemented through m1 data type (\texttt{m8->m1->m4}) according to the compiler features).

\section{Numerical Results}

\subsection{Hardware}

The experiments are carried out using the following software and hardware environment. The Lobachevsky supercomputer node in the following configuration is considered as a basic device for comparison: 2~x~\textbf{Intel Xeon Silver 4310T} CPU 2.30GHz, 64 GB RAM. Only one processor (10 cores) is used to avoid NUMA effects, which are not present in RISC-V devices. Threads are bound to computational cores using affinity masks. The CentOS~7 operating system and GCC~9.5 compiler are employed. 

Other performance data are collected on two RISC-V devices supporting vector instruction sets. The first device is \textbf{Mango Pi MQ-Pro (D1)} with an Allwinner D1 processor (1~x~XuanTie C906, 1GHz) and 1GB DDR3L RAM. Given the rapid progress in the development of RISC-V devices, this board is already outdated. However, by comparing its performance to more modern CPUs, we can estimate the pace of development in this area. Below are some characteristics of Mango Pi. The device contains only one computational core that supports the RV64IMAFDCV instruction set, a five-stage pipeline, a single-level 2-associative L1 cache of 32 KB in size and a cache line size of 64 bytes, an associative translation buffer, branch prediction, hardware prefetching for instructions and data, 16, 32, and 64-bit integer and floating point operations, as well as 128-bit vector operations, including, importantly for many applications, the FMA floating point operation (fused multiply-add: $d=a+bc$). The Ubuntu~22.10 operating system (RISC-V edition) and GCC~12.2 compiler are installed.

\textbf{Sipeed Lichee Pi 4A} is the second and more modern RISC-V device. This is a single-board computer (SBC) built on the Alibaba T-Head TH1520 platform. In addition to the Xuantie C910 processor with four RISC-V cores (RV64GCV, clocked up to 2.5 GHz, our sample runs at 1.85 GHz) and 8 GB of LDDR4X RAM, the system includes a neuroprocessing unit (NPU) with a performance of up to 4 TOPS@INT8, power efficient Xuantie E902 core, Imagination 3D graphics unit (50 Gflops) and Xuantie C906 DSP. The developers of Lichee Pi 4A position it as an analog of Raspberry Pi 4 in terms of performance and applications \cite{c21}. The Xuantie C910 CPU includes four RV64GC architecture cores supporting 64-bit RISC-V ISA (the symbol ``G'' in the acronym), 16-bit compact instructions (``C''), 32-bit regular instructions, 16, 32, and 64-bit floating point operations, including FMA, vector instructions RVV~0.7.1, and more than 50 custom instructions to speed up various algorithms (XuanTie Instruction Extension, XIE). The superscalar pipeline consists of 12 stages and supports decoding from 3 to 8 instructions per cycle, containing instruction prefetch and branch prediction algorithms. The L1I and L1D caches of each core have a size of 64 KB, a cache line size of 64 bytes and an associativity of 2, a total inclusive L2 cache of 1 MB has an associativity of 16 and supports data prefetching. The MESI and MOESI coherence protocols are implemented for the L1 and L2 caches, respectively. The memory management module complies with the RISC-V SV39 standard and uses 2 levels of TLB cache \cite{c22,c23}.

\subsection{Benchmarking Methodology}

This project aims to accelerate OpenCV algorithms through enhanced vectorized computations. Therefore, the experimental part of the project is as follows. We select several OpenCV algorithms so that vectorization on the x86 CPU reduces computation time compared to a scalar version. For these algorithms, we run the following versions of the code:

\begin{enumerate}
  \item Sequential scalar version (\textit{``SeqScalar''}). 
  \item Parallel scalar version (\textit{``ParScalar''}). 
  \item Sequential vectorized version (\textit{``SeqVector''}). 
  \item Parallel vectorized version (\textit{``ParVector''}). 
  \item Optimized parallel vectorized version (\textit{``Optim''}, improved vectorization proposed in this paper). 

\end{enumerate}

We run these implementations on the x86 server and two RISC-V CPUs and check correctness. The computation time is measured in several experiments with the selection of the shortest time. We only use parallel versions for pre-parallelized OpenCV algorithms.

\subsection{Image Filtering Algorithm: Benchmarking and Performance Analysis}

The first algorithm solves the classical problem of image processing -- filtering images using the Gaussian filter kernel. The filtering problem is as follows. Let there be an image containing one or three channels at the input. Each image pixel contains one or three intensity values, respectively, each ranging from 0 to 255, or 0 to 1 if pre-normalized. The filtering task includes passing through the image from left to right and from top to bottom, applying the Gaussian filter kernel to the corresponding pixels, and calculating a two-dimensional discrete convolution. The result of the algorithm is an image that has the same spatial dimensions as the input image and contains updated intensity values \cite{c6}.

\begin{table}[ht]
\caption{Performance results of image filtering on the x86 CPU. Time is given in seconds}
\label{tab:1}
\begin{tabular}{|c|c|c|c|c|}
\hline
Resolution & Kernel size & SeqScalar & SeqVector & \begin{tabular}[c]{@{}c@{}}Vectorization \\ speedup\end{tabular}  \\ \hline
1920x1080 & 3x3 & 0,058 & 0,010 & 6,06 \\
1920x1080 & 5x5 & 0,168 & 0,022 & 7,70 \\
1920x1080 & 7x7 & 0,376 & 0,040 & 9,35 \\
1920x1080 & 9x9 & 0,641 & 0,065 & 9,89 \\
1920x1080 & 11x11 & 1,071 & 0,095 & 11,24 \\
1920x1080 & 13x13 & 0,177 & 0,152 & 1,17 \\ \hline
3840x2160 & 3x3 & 0,234 & 0,039 & 6,06 \\
3840x2160 & 5x5 & 0,674 & 0,088 & 7,69 \\
3840x2160 & 7x7 & 1,503 & 0,161 & 9,37 \\
3840x2160 & 9x9 & 1,740 & 0,259 & 6,72 \\
3840x2160 & 11x11 & 2,904 & 0,381 & 7,62 \\
384x2160 & 13x13 & 0,673 & 0,574 & 1,17 \\ \hline
\end{tabular}
\end{table}

Table \ref{tab:1} shows the results of running the \texttt{filter2D()} function from the OpenCV library on the x86 CPU. Given that the implementation in OpenCV was not parallelized, we only run the serial version. It can be seen that vectorization significantly speeds up the code, but the speedup fluctuates from 1.17 to 11.24. A possible reason for the lack of significant acceleration for large kernels is that they use a different algorithm based on the Fast Fourier Transform. 

\begin{table}[ht]
\caption{Performance results of image filtering on Mango Pi. Time is given in seconds}
\label{tab:2}
\begin{tabular}{|c|c|c|c|c|c|c|}
\hline
Resolution & Kernel size & SeqScalar & SeqVector & \textbf{Optim} & \begin{tabular}[c]{@{}c@{}}Vectorization \\ speedup\end{tabular} & \textbf{\begin{tabular}[c]{@{}c@{}}Optimization\\ speedup\end{tabular}} \\ \hline
1920x1080 & 3x3 & 1,26 & 0,69 & \textbf{0,76} & 1,84 & \textbf{0,90} \\
1920x1080 & 5x5 & 2,61 & 1,42 & \textbf{1,79} & 1,84 & \textbf{0,80} \\
1920x1080 & 7x7 & 4,34 & 2,72 & \textbf{3,29} & 1,59 & \textbf{0,83} \\
1920x1080 & 9x9 & 6,52 & 6,00 & \textbf{5,97} & 1,09 & \textbf{1,01} \\
1920x1080 & 11x11 & 6,55 & 6,26 & \textbf{5,98} & 1,05 & \textbf{1,05} \\
1920x1080 & 13x13 & 6,55 & 6,08 & \textbf{6,00} & 1,08 & \textbf{1,01} \\ \hline
3840x2160 & 3x3 & 5,42 & 2,97 & \textbf{3,27} & 1,83 & \textbf{0,91} \\
3840x2160 & 5x5 & 10,53 & 6,15 & \textbf{7,38} & 1,71 & \textbf{0,83} \\
3840x2160 & 7x7 & 17,97 & 10,99 & \textbf{13,48} & 1,63 & \textbf{0,82} \\
3840x2160 & 9x9 & 24,48 & 22,34 & \textbf{22,20} & 1,10 & \textbf{1,01} \\
3840x2160 & 11x11 & 24,43 & 22,62 & \textbf{22,38} & 1,08 & \textbf{1,01} \\
384x2160 & 13x13 & 25,10 & 23,33 & \textbf{22,48} & 1,08 & \textbf{1,04} \\ \hline
\end{tabular}
\end{table}

\begin{table}[ht]
\caption{Performance results of image filtering on Lichee Pi. Time is given in seconds}
\label{tab:3}
\begin{tabular}{|c|c|c|c|c|c|c|}
\hline
Resolution & Kernel size & SeqScalar & SeqVector & \textbf{Optim} & \begin{tabular}[c]{@{}c@{}}Vectorization \\ speedup\end{tabular} & \textbf{\begin{tabular}[c]{@{}c@{}}Optimization\\ speedup\end{tabular}} \\ \hline
1920x1080 & 3x3 & 0,19 & 0,20 & \textbf{0,14} & 0,97 & \textbf{1,41} \\
1920x1080 & 5x5 & 0,31 & 0,44 & \textbf{0,44} & 0,71 & \textbf{1,00} \\
1920x1080 & 7x7 & 0,48 & 0,78 & \textbf{0,71} & 0,62 & \textbf{1,11} \\
1920x1080 & 9x9 & 0,61 & 0,64 & \textbf{0,59} & 0,95 & \textbf{1,08} \\
1920x1080 & 11x11 & 0,61 & 0,65 & \textbf{0,60} & 0,94 & \textbf{1,09} \\
1920x1080 & 13x13 & 0,61 & 0,65 & \textbf{0,60} & 0,95 & \textbf{1,09} \\ \hline
3840x2160 & 3x3 & 0,79 & 0,80 & \textbf{0,58} & 0,99 & \textbf{1,39} \\
3840x2160 & 5x5 & 1,25 & 1,80 & \textbf{1,39} & 0,69 & \textbf{1,29} \\
3840x2160 & 7x7 & 1,92 & 3,12 & \textbf{2,79} & 0,61 & \textbf{1,12} \\
3840x2160 & 9x9 & 2,33 & 2,45 & \textbf{2,23} & 0,95 & \textbf{1,10} \\
3840x2160 & 11x11 & 2,34 & 2,46 & \textbf{2,24} & 0,95 & \textbf{1,10} \\
3840x2160 & 13x13 & 2,33 & 2,46 & \textbf{2,26} & 0,95 & \textbf{1,09} \\ \hline
\end{tabular}
\end{table}

Let's examine the results on Mango Pi (Table \ref{tab:2}) and Lichee Pi (Table \ref{tab:3}) RISC-V devices. First, we note that conventional vectorization, unlike x86, leads to no more than a twofold speedup. This is due, among other things, to the fact that 128-bit registers are used against 512-bit ones in the x86 CPU, and not in the most optimal way. On Mango Pi the developed optimization does not lead to a significant gain, and in some examples leads to a slowdown. However, when switching to a more modern Lichee Pi, the results are much better. ``Standard'' vectorization does not lead to speedup, however, the optimization performed using ``wide'' registers improves performance from 8\% to 41\% depending on the size of the image and the kernel. Also note the approximately tenfold speedup compared to the previous generation of the device (Mango Pi) for large images, which shows the pace of progress in the development of RISC-V processors. Compared to the x86 CPU, the largest image only shows a fourfold slowdown, but keep in mind that we are comparing vectorized, but sequential versions of the code. In the case of parallelization of the algorithm, the gap, of course, will increase in accordance with the difference in the number of computational cores.

\begin{table}[ht]
\caption{Performance results of the erosion algorithm on the x86 CPU. Time is given in seconds}
\label{tab:4}
\begin{tabular}{|c|c|c|c|c|}
\hline
Resolution & Filter size & SeqScalar & SeqVector & \textbf{\begin{tabular}[c]{@{}c@{}}Vectorization \\ speedup\end{tabular}}  \\ \hline
1920x1080 & 1 & 0,007 & 0,001 & \textbf{5,17} \\
1920x1080 & 2 & 0,010 & 0,001 & \textbf{6,78} \\
1920x1080 & 3 & 0,013 & 0,001 & \textbf{8,60} \\ \hline
3840x2160 & 1 & 0,025 & 0,005 & \textbf{5,59} \\
3840x2160 & 2 & 0,037 & 0,005 & \textbf{7,61} \\
3840x2160 & 3 & 0,050 & 0,005 & \textbf{9,92} \\ \hline
7680x4320 & 1 & 0,099 & 0,014 & \textbf{6,92} \\
7680x4320 & 2 & 0,098 & 0,011 & \textbf{9,07} \\
7680x4320 & 3 & 0,134 & 0,012 & \textbf{11,14} \\ \hline
15260x8640 & 1 & 0,259 & 0,041 & \textbf{6,29} \\
15260x8640 & 2 & 0,391 & 0,045 & \textbf{8,66} \\
15260x8640 & 3 & 0,532 & 0,049 & \textbf{10,76} \\ \hline
\end{tabular}
\end{table}

\subsection{Erosion Algorithm: Benchmarking and Performance Analysis}

The second algorithm considered in this paper performs ``erosion'', which is also one of the basic image processing operations. The essence of erosion is to bypass a single-channel image with a kernel of a certain size (kernel size is an algorithm parameter) and combine the anchor of the kernel with the current pixel of the image, by analogy with the filtering operation. The only difference is that instead of calculating the operation of a two-dimensional discrete convolution, a search is made for the minimum intensity in the neighborhood covered by the kernel. The intensity value of the pixel corresponding to the anchor is replaced by the minimum in its vicinity, as a result of which the white areas are narrowed in the image.

\begin{table}[ht]
\caption{Performance results of the erosion algorithm on Mango Pi. Time is given in seconds}
\label{tab:5}
\begin{tabular}{|c|c|c|c|c|c|c|}
\hline
Resolution & Filter size & SeqScalar & SeqVector & Optim & \textbf{\begin{tabular}[c]{@{}c@{}}Vectorization \\ speedup\end{tabular}} & \textbf{\begin{tabular}[c]{@{}c@{}}Optimization\\ speedup\end{tabular}} \\ \hline
1920x1080 & 1 & 0,163 & 0,060 & 0,043 & \textbf{2,71} & \textbf{1,40} \\
1920x1080 & 2 & 0,246 & 0,073 & 0,048 & \textbf{3,37} & \textbf{1,51} \\
1920x1080 & 3 & 0,320 & 0,084 & 0,054 & \textbf{3,79} & \textbf{1,57} \\ \hline
3840x2160 & 1 & 0,586 & 0,242 & 0,157 & \textbf{2,42} & \textbf{1,54} \\
3840x2160 & 2 & 0,900 & 0,297 & 0,180 & \textbf{3,03} & \textbf{1,65} \\
3840x2160 & 3 & 1,189 & 0,361 & 0,201 & \textbf{3,29} & \textbf{1,79} \\ \hline
7680x4320 & 1 & 2,504 & 0,981 & 0,605 & \textbf{2,55} & \textbf{1,62} \\
7680x4320 & 2 & 3,925 & 1,218 & 0,845 & \textbf{3,22} & \textbf{1,44} \\
7680x4320 & 3 & 5,178 & 1,460 & 0,780 & \textbf{3,55} & \textbf{1,87} \\ \hline
15260x8640 & 1 & 12,499 & 11,421 & 9,923 & \textbf{1,09} & \textbf{1,15} \\
15260x8640 & 2 & 24,458 & 12,457 & 10,404 & \textbf{1,96} & \textbf{1,20} \\
15260x8640 & 3 & 29,452 & 13,726 & 10,832 & \textbf{2,15} & \textbf{1,27} \\ \hline
\end{tabular}
\end{table}

\begin{table}[ht]
\caption{Performance results of the erosion algorithm on Lichee Pi. Time is given in seconds}
\label{tab:6}
\begin{tabular}{|c|c|c|c|c|c|c|}
\hline
Resolution & Filter size & SeqScalar & SeqVector & Optim & \textbf{\begin{tabular}[c]{@{}c@{}}Vectorization \\ speedup\end{tabular}} & \textbf{\begin{tabular}[c]{@{}c@{}}Optimization\\ speedup\end{tabular}} \\ \hline
1920x1080 & 1 & 0,020 & 0,004 & 0,004 & \textbf{4,56} & \textbf{1,17} \\
1920x1080 & 2 & 0,031 & 0,005 & 0,004 & \textbf{5,91} & \textbf{1,27} \\
1920x1080 & 3 & 0,049 & 0,006 & 0,004 & \textbf{7,93} & \textbf{1,37} \\ \hline
3840x2160 & 1 & 0,060 & 0,017 & 0,015 & \textbf{3,47} & \textbf{1,17} \\
3840x2160 & 2 & 0,095 & 0,023 & 0,016 & \textbf{4,21} & \textbf{1,39} \\
3840x2160 & 3 & 0,147 & 0,026 & 0,018 & \textbf{5,56} & \textbf{1,44} \\ \hline
7680x4320 & 1 & 0,235 & 0,071 & 0,058 & \textbf{3,29} & \textbf{1,23} \\
7680x4320 & 2 & 0,379 & 0,087 & 0,068 & \textbf{4,34} & \textbf{1,29} \\
7680x4320 & 3 & 0,590 & 0,104 & 0,078 & \textbf{5,68} & \textbf{1,33} \\ \hline
15260x8640 & 1 & 0,931 & 0,307 & 0,283 & \textbf{3,03} & \textbf{1,09} \\
15260x8640 & 2 & 1,530 & 0,370 & 0,317 & \textbf{4,13} & \textbf{1,17} \\
15260x8640 & 3 & 2,363 & 0,439 & 0,360 & \textbf{5,38} & \textbf{1,22} \\ \hline
\end{tabular}
\end{table}

We run the erosion algorithm implemented in the OpenCV \texttt{erode()} function on three target devices. The results of runs on the x86 CPU are shown in Table \ref{tab:4}. All runs were performed in serial mode, since this algorithm is not parallelized in OpenCV, but we are primarily interested in vectorization. We note a good acceleration from vectorization (up to $\sim 11$ times).

Consider the performance results on Mango Pi (Table \ref{tab:5}) and Lichee Pi (Table \ref{tab:6}). Experiments on Mango Pi show that the basic version of the vectorized code implemented in the OpenCV library using universal intrinsics generally shows a good speedup of up to 3.79 times. However, this result can be greatly improved by applying the approach proposed in this paper. Additional speedup compared to the basic vectorized version is from 1.15 to 1.87 times. When switching to a much faster Lichee Pi, the calculation time is reduced by more than an order of magnitude, while the code is still vectorized better than on Mango Pi, and the developed optimization allows for an additional acceleration up to 1.44 times.

\subsection{Bag-of-words and Support Vector Machine Algorithms: Benchmarking and Performance Analysis}

Finally, we perform experiments with the bag-of-words and the Support Vector Machine (SVM) algorithms. SVM is one of the commonly used ML algorithms. An image classification problem is considered as a testbed. In this regard we employ the Cifar-10 dataset \cite{c17}, which contains 50,000 training and 10,000 test images with a resolution of 32x32 pixels. To extract features from an image, an approach called "bag-of-words" is used. Any ML algorithm works in two stages: training and testing the model. 

Training involves the following steps \cite{c18-1,c18-2}:

\begin{enumerate}
  \item Detecting of keypoints on each image of the training dataset. In our experiments, we use SIFT \cite{c19} as a detection algorithm. 
  \item Constructing of descriptors of keypoints. SIFT \cite{c19} is further used to construct descriptors. 
  \item Clustering of descriptors of keypoints belonging to all objects of the training set. As a result of clustering, the construction of a dictionary is ensured, the ``words'' in which are the centroids of the constructed clusters. 
  \item Building a normalized histogram of the occurrence of ``words'' for each image of the training sample. For each cluster, the number of keypoints assigned to it and belonging to a certain image is calculated. 
  \item Training of a classifier using the feature description of the image calculated at step 4. In our case, SVM is used as the classification algorithm.
\end{enumerate}

Testing assumes the presence of a trained SVM model and a dictionary built at step 3 of the training stage and involves the following steps:
\begin{enumerate}
  \item Detection of keypoints on each image of the test dataset. In this case, it is necessary to use the same algorithm for detecting keypoints as at the training stage. 
  \item Building a normalized histogram of the occurrence of ``words'' for each image of the test sample. The dictionary is obtained at the training stage. The construction of descriptors of keypoints is carried out similarly to the previous step using the same algorithm as for training. 
  \item Image class prediction using the trained SVM model.

\end{enumerate}

Classifier training is a procedure that is usually performed offline on high-performance computing devices, while testing is carried out repeatedly on the target hardware, which may consist of cheap low-performance devices. Therefore, the optimization of the testing stage is of interest. In our experiments we measure computation time of all three stages of the testing algorithm, namely (I) keypoint detection, (II) feature generation, (III) prediction. 

Consider the results on the x86 server (table \ref{tab:7}). We perform experiments for all types of SVM kernels implemented in OpenCV, but main observations do not depend on the type of the kernel because only the time of the third stage increases, but not the acceleration from parallelization or vectorization. Therefore, we present the results only for the linear kernel. The dictionary size was chosen to be 250. First, it should be noted that there is no significant speedup from parallelization in the implementations from OpenCV. There is a rather small acceleration only at the stage of feature extraction. However, vectorization accelerates the first two stages by two or three times, while the third stage is not vectorized.

\begin{table}[ht]
\caption{Performance results of the SVM algorithm on the x86 CPU. Time is given in seconds}
\label{tab:7}
\begin{tabular}{|c|c|c|c|c|c|}
\hline
SVM step & SeqScalar & ParScalar & SeqVector & ParVector & \textbf{\begin{tabular}[c]{@{}c@{}}Vectorization \\ speedup\end{tabular}}  \\ \hline
keypoint detection & 9,21 & 9,07 & 3,54 & 3,84 & \textbf{2,60} \\
feature generation & 10,85 & 8,63 & 5,10 & 3,67 & \textbf{2,36} \\
prediction & 0,06 & 0,06 & 0,06 & 0,06 & \textbf{0,99} \\ \hline
\end{tabular}
\end{table}

\begin{table}[ht]
\caption{Performance results of the SVM algorithm on Mango Pi. Time is given in seconds}
\label{tab:8}
\begin{tabular}{|c|c|c|c|c|c|}
\hline
SVM step & SeqScalar & SeqVector & Optim & \textbf{\begin{tabular}[c]{@{}c@{}}Vectorization \\ speedup\end{tabular}} & \textbf{\begin{tabular}[c]{@{}c@{}}Optimization\\ speedup\end{tabular}} \\ \hline
keypoint detection & 269,67 & 231,86 & 194,38 & \textbf{1,16} & \textbf{1,19} \\
feature generation & 388,95 & 338,99 & 298,82 & \textbf{1,15} & \textbf{1,13} \\
prediction & 3,27 & 3,25 & 3,24 & \textbf{1,01} & \textbf{1,00} \\ \hline
\end{tabular}
\end{table}

\begin{table}[ht]
\caption{Performance results of the SVM algorithm on Lichee Pi. Time is given in seconds}
\label{tab:9}
\begin{tabular}{|c|c|c|c|c|c|c|c|}
\hline
SVM step & SeqScalar & ParScalar & SeqVector & ParVector & Optim & \textbf{\begin{tabular}[c]{@{}c@{}}Vectorization \\ speedup\end{tabular}} & \textbf{\begin{tabular}[c]{@{}c@{}}Optimization\\ speedup\end{tabular}} \\ \hline
\begin{tabular}[c]{@{}c@{}}keypoint \\    detection\end{tabular} & 26,82 & 24,59 & 25,71 & 24,16 & 19,27 & \textbf{1,02} & \textbf{1,25} \\
\begin{tabular}[c]{@{}c@{}}feature \\    generation\end{tabular} & 36,02 & 25,96 & 36,71 & 26,76 & 21,30 & \textbf{0,97} & \textbf{1,26} \\
prediction & 0,30 & 0,30 & 0,30 & 0,30 & 0,30 & \textbf{1,00} & \textbf{1,00} \\ \hline
\end{tabular}
\end{table}

Consider the results of experiments on Mango Pi (Table \ref{tab:8}) and Lichee Pi (Table \ref{tab:9}). Regarding parallelization, they are generally similar to x86: only a slight speedup on Lichee Pi is observed at the second stage of the method. The standard vectorization from OpenCV on Mango Pi speeds up the first two steps, while the reference version does not speed up on Lichee Pi. The developed optimization also gives a gain on both devices of the RISC-V architecture from 1.13 to 1.26, depending on the device and the stage of the algorithm.

\section{Conclusion}
The paper examines the performance of available prototypes of RISC-V devices using three basic machine learning and computer vision algorithms implemented in the widely used OpenCV library. The results show that the library can be easily used on RISC-V devices. It is enough just to rebuild the code with the appropriate compiler. When running algorithms on RISC-V processors, correct results are obtained. The performance is inferior to the x86 server CPU, but when moving from Mango Pi to the next edition of the device (Lichee Pi) the computation time is often reduced by an order of magnitude, which indicates rapid progress in this field. Probably, we can expect the development of a competitive RISC-V CPUs suitable for high-performance computing.

The main contribution of the paper is the optimization of vector implementations of OpenCV algorithms, which improves performance by several tens of percent. The optimization is based on taking into account the peculiarities of the vector extensions of the RISC-V instruction set and allows for more efficient use of register blocks compared to the conventional implementation. Note that this improvement was quickly integrated into the OpenCV code thanks to the use of the concept of ``universal intrinsics'' in the library, abstracting the code from a specific architecture. We plan to continue work in the field of improving performance of various OpenCV algorithms for existing and future RISC-V CPUs.

%\begin{acknowledgments}
%The project is supported by the Lobachevsky University Academic Leadership Program ``Priority-2030''
%\end{acknowledgments}

%
% The Bibliography
%

\end{document}